\documentclass{nature}

\usepackage{graphicx}% Include figure files
\usepackage{color}% bold math
\title{Disorder driven inhomogeneous phase in the 2D-superconducting film of titanium nitride}

\author{P. Kulkarni$^{1*}$, S. Vieira$^{1}$, J. Gabriel$^{1}$, M. Baklanov$^{2}$, T. Baturina$^{3}$, V. Vinokur$^{4}$}

\begin{document}

\maketitle

\begin{affiliations}

\item Laboratorio de Bajas Temperaturas, Departamento de F\'isica de la Materia Condensada, Instituto de Ciencia de Materiales Nicol\'as Cabrera, Facultad de Ciencias, Universidad Aut\'onoma de Madrid, E-28049 Madrid, Spain
\item{IMEC, Kapeldreef 75, B-3001 Leuven, Belgium}
\item{Institute of Semiconductor Physics, 13 Lavrentjev Avenue, Novosibirsk, 630090, Russia}
\item{Materials Science Division, Argonne National Laboratory, Argonne, Illinois 60439, USA}

 %\item separate with \verb|\item| commands.
\end{affiliations}

\begin{abstract}

Typically the superconducting phase weakens at several points with the increase in disorder before it is distroyed in the 2d-thin films. This may lead to an inhomogeneous superconducting state without a continuous phase. Here we present scanning tunneling spectroscopy measurements at 0.1 K in the disordered polycrystalline film of TiN describing the nanoscale size features of the superconducting state. The imaging shows imcommensurate charge density modulations, originating at the crystalline bounadries, and intercepted on large scale by the beat patterns in the regions of overlap. Electronic coherence is maintained over length scale minimum of crystalline sizes, and suffers scattering across low angle crystalline boundaries. The superconducting state fluctuates at the positions of the charge density modulations and zones of the weak phase appear in the vicinity of the beats. Our data shows that the BCS-like behavior evolves into the V-shaped density of states in such inhomogeneous regions as a result of the competition between the superconducting correlations with that of the strong electron-electron repulsive interactions assisted by the inelastic scattering at the crystalline boundaries.

\end{abstract}

Two dimensional thin films of superconducting materials are interesting, because the divergence of their resistance curve at the transition may vary from one extreme to another at low temperatures with the increase in the magnetic field, disorder strength or charge density. The quantum of resistance $h/4e^2$ governs the crossover situation where a superconducting behavior is lost and electron localization sets in \cite{Strongin70,Haviland,Hebard,Fisher,Baturina,Baturina1}. To explain the microscopic drive of this quantum phase transition (studied experimentally at finite temperatures) the discussion revolves around Josephson coupling between grains and its competition with Coulomb interactions\cite{Imry81}. Popular approaches in theoretical and transport measurements being to perceive the changes in the superconducting order parameter locally close to the superconductor to insulator transition (SIT). The local tunneling measurements to a large extent supported the fluctuations of the superconducting order parameter. However, these measurements till now were not based on the high resolution simultaneous topographic information.

Many interesting local tunneling experiments addressed the effect of disorder on the superconducting phase. Intrinsic inhomogeneities were reported as a function of amplitude fluctuations in strongly disordered TiN films\cite{Sacepe} or phase fluctuations in NbN films\cite{Noat}. In homogeneously disordered InO films, STS measurements of the tunneling spectra were reported close to SIT\cite{Sacepe1}. It was shown that the quasi-particle peaks tend to suppress with increase in disorder while the superconducting gap remains intact\cite{Sacepe1}. Direct localization of the preformed cooper pairs was suggested as the precursor to the SIT. On the other hand the evolution of the tunneling spectra with temperature in TiN films suggested the pseudo-gap like behavior\cite{Sacepe2}, which was considered as a precursor to the superconducting phase. Pseudo-gap features were also shown to enhance with the increase in the disorder. These studies were aimed at observing the local changes in the superconducting phase at different levels of disorder. However, the role of Coulomb interactions remained obscure in local measurements. We chose well characterized TiN films\cite{Baturina,Baturina1,Pfuner} to study the atomic level topographic and high resolution spectroscopic images using STM/S at 100 mK. Our measurements reveal possibilities to observe the phase break at the spatial positions leading to the inhomogeneous superconducting state due to the role of Coulomb interactions in the disordered films.

The atomic scale STM experiment is, in principle, a rather complex one. The tunneling current is the result of the quantum mechanical coupling between sample and tip wavefunction's at a given position, which does not only depend on the precise electronic structure of the surface, but also on the atomic orbitals of the tip and the conduction through them. However, low energy scale phenomena, featuring the current at low bias voltages, can be interpreted most easily\cite{Tersoff85}. It is shown that the tip nearly always effectively adopts d-wave form orbitals which interact with the underlying atomic orbitals to produce sharp images of the charge distribution. Thus, images show the charge distribution around single atoms, and eventually, charge fluctutions due to additional electronic ordering, whose magnitude is generally incommensurate to atomic modulations. The single atom features give Bragg peaks at the atomic positions, and the charge fluctuations at fixed points of the Fermi surface at which scattering has lead to the development of a standing wave. On the other hand, the tunneling density of states gives precise information about pairing magnitude and its dependence as a function of the position and magnetic field, at atomic scale.

In this manuscript we explore the nanoscale superconducting features in disordered ultrathin film of TiN using a homemade scanning tunneling microscopy and spectroscopy (STM/S) set up in a dilution refrigerator.The titanium nitride film was deposited using Atomic layer chemical vapour deposition technique at 350 degree on Si/SiO$_2$ substrate. Details about the experiment and sample preparation are given in the Supplementary Information (SI). We measured a superconducting sample with a critical temperature of 1 K, B$_{c2}(0)=2.5$ T and $\xi=9.3$ nm.

\paragraph{Nano scale scanning tunneling microscopy in the disordered 2d-TiN film\\}

The film thickness was 5 nm and the microstructure consists of densly packed crystallites with low angle boundaries. The average crystallite size range was estimated as 3 to 8 nm, with the mean size around 5 nm. TiN has FCC structure with the atomic distance between (100) planes as 4.2 $\AA$. In Figure 1 (a), the positions of Ti and N ions are shown in the unit cell. The STM image adjacent in panel (b) shows a representative morphology of the TiN film surface, where several nanocrystallites are visible in the contrast and the histogram in the inset shows their size distribution. Crystallites are not regular shaped, and therefore lateral sizes are often in the ratio from 1 to 3. Lower panels (c) and (d) in Figure 1 show the atomic level imaging over an area 10 nm x 10 nm. The atmoic positions form regular arrays in two directions. The height variations in the image are less than 0.5 nm. A 2d-FFT of the atomic level image is shown in panel (d), where it is easier to locate the periodic structures. A simple $k$ = 1/wavelength formula is used to make the 2d-FFT images presented in this paper. In panel (d), several bright spots are seen, regularly arranged forming replicas of two dimensional atomic lattice in panel (c). The distances of the first set of bright spots from the centre are measured and converted in the units of lattice parameter $a$ for TiN. As highlighted in the panel (d), these would correspond to [200] and [110] sets of atomic planes in FCC structure of TiN. Eight panels (e-l) on the right are the STM images on larger scales and their 2d-FFTs. Panels (e), (g), (i), (k) show the topographic images with four different scan sizes in the decreasing order. In panel (e) over a large scale image with an area about 512 nm x 512 nm, several bright lines are observed. With small area scans, 56.9 nm x 56.9 nm in panel (g) and 42.7 nm x 42.7 nm in panel (i), these lines are resolved as charge density modulations coherent over distances several 10s of nm and overlapping considerably forming beat patterns. The intensity at these overlapped regions becomes stronger at the beating positions in the large area scan in panel (e), and the beats are seen as bright lines. In the panel (f), the 2d-FFT image shows (along the line profile between the white arrows) the k-vector with several magnitudes, which falls expontially as seen in the inset. This shows the long range interactions of the electronic charge density modulations. When the scan size is reduced in panels (g) and (i) the respective 2d-FFT images in panels (h) and (j), the bright spots are seen forming rectangular positions, corresponding to the charge density modulations. In panel (k) in Figure 1, the scan image is shown over 12.7 nm x 12.7 nm. Two orienations of the charge modulations can be seen along with the atomic level features in the topographic image. A blue and green color lines identify these two orienations. In panel (l), we show the 2d-FFT image of panel (k) along with the magnified portion in the inset. We observe four bright spots, for two orienations of the charge modulations, encirled with green and blue lines, respectively. At smaller wavelengths, as shown in the main panel (l), six bright spots are circled using black lines, marking the underlying hexagonal atomic positions in the image in panel (k). The distances of these bright spots in the units of lattice parameter $a$ would form the planes [210] and [200]. The lattice of TiN in the atomic images results from the small single crystallites oriented randomly where the exposed surface is on the the cuts of the FCC lattice, giving atomic size features of either rectangular, square or hexagonal symmetry. The observed modulation wavevectors can be associated to one of the principal directions of the bulk FCC lattice, implying that the growth is produced in the same way as in the bulk,  however, along the strongly differing orientations. Average crystallite size is between 3-8 nm, and the atomic lattices are separated by the low angle crystalline boundaries. The disordered TiN film shows clearly additional coherent charge density modulations incommensurate with the atomic lattice (also see Figure S2). These modulations have wavevectors between 2-10 times the interatomic distances. These modulations with sizes between 20-40 nm cross several crystallite boundaries. This is a purely electronic effect, showing vividly electron localization in the form of standing waves, and producing an unexpected charge ordering pattern, superposed to the crystalline structure. These images demonstrate the existence of "coherence patches", whose size is orders of magnitude above the mean free path estimated from resistivity to about 10 nm. These patches will conform to a large extent the macroscopic electronic conduction behavior. Let us review the known electronic properties of TiN. The Fermi surface and bandstructure of bulk TiN has been obtained in de Haal van Alphen and angular resolved photoemission experients. In the bulk, there are three bands crossing the Fermi surface, two of them giving small nearly cubic Fermi surfaces close to the zone center, and one rather involved large three dimensional Fermi surface. Fermi wavevectors are distributed over the whole Brillouin zone. As is well known, STM can probe, in addition to atomic arrangements, charge dentisy waves. These may be either an intrinsic deformation of the electronic bandstructure, in which case they have a well defined orientation and wavevector over the whole sample, or they may appear due to inelastic electron scattering at the crystalline boundries. As we are measuring a very thin film, with a thickness well below the length scale over which modulations extend, we are probing the charge pattern formed by disorder in the whole film. Generally, electronic waves are formed from scattering events occuring between two parts of the Fermi surface, joined by a Fermi wavevector $\lambda$$_F$ which gives the wavelength of the charge modulations in the STM image. We can compare the $\lambda$$_F$ measured from our images with the length scales characteristic of the TiN metal obtained from macroscopic resistance measurements. Taking the diffusion constant, of $D$ = 0.32 cm$^2$/s, and $\tau$$_F$ = 10$^{-15}$ s, obtained from the 2D Fermi energy expression, E$_F$ = $\pi$$h$$^{2}$$n/4m$ with electron density n = 5 x 10$^{21}$ m$^{-3}$ we find $\lambda$$_F$ = 1 nm, of order of the size of the charge modulations we observed in the images (Figure S2). As shown in the patterns in large scale images, charge density modulations are observed particularly along preferential directions. This shows that scattering is between small parts of the Fermi surface at an angle to the high symmetry axis.

\paragraph{STS mapping in the superconducting state in 2d-disordered TiN film\\}

With a reasonable understanding of the sample under scanner, we proceed with the studies of the local electronic density of states. A STS map was made at 100 mK in the bias voltage range from -1 mV to +1 mV on the the area in Figure 1(e). The STS map comprises measuring 128 x 128 IV tunneling curves and numerically derivating to plot the conductance curves normalized at 1 mV. The spatial changes in the tunneling conductance values can be traced at a fixed bias voltage. Representative tunneling conductance curves at two positions are shown in Figure 2 (a) in the same color scale as in the spectromap in panel (b). The bias voltage of -0.25 mV was chosen to display the distribution of the tunneling conductance spatially. Clearly the conductance changes at the selected bias voltage highlights the modulations of the density of states. At the positions of the beats the shape of the superconducting density of states tend to fill up close to the Fermi level, and the height of the quasi-particle peak positions tend to suppress proportionally. In panel (c) the average tunneling conductance for each curve between -1 to +1 mV was calculated, and plotted spatially over the same area. The map consists of total 128 x 128 values of the average conductance obtained from each tunneling conductance curve. Inset in panel (c) shows the histogram of the average of the sum of the conductance, which peaks around 0.7 over the scale of 0 to unity. In the main panel (c) the color scale is same as that in the panel (b), and clearly the average conductance has much sharper width than the conductance values at a specific bias voltage. This points to the fact that the total density of states remain same, while it is distributed in spatially varying behavior in the superconducting state. We extend the STS measurement to other larger scan areas. The tunneling conductance behavior changes spatially as shown in the inset in the panel Fig.2 (d). The area under investigation was 455 nm x 455 nm and for the most part of it the superconducting gap opens with the BCS-like quasi-particle peaks (inset in panel (d)), however, in several curves the peaks are either suppressed or altogether absent, with the equivalent rounding near zero bias voltage. In order to generate a picture of the conductance distribution at the selected bias voltage of 0.25 mV, all the tunneling conductance curves are normalized at 1 mV. The normalized conductance varies from 0 to 1, as shown in the color scale in Fig. 2(d). The dark blue color is the region where the conductance at 0.25 mV reaches near zero values, while the red region has the conductance close to 0.7 and is also a representative of the region where the curves without quasi-particle peaks are seen. Spatial changes were reported previously in terms of the average superconducting gap of Ref.[8], however a clear distinction between two behaviors, with and without quasi-particle peaks, was not established. In the panel (e), the average of the sum of the conductance in the bias voltage range -1 to +1 mV (on the color scale from 0 to 1) for each of the 128 x 128 curves is plotted over the same area. The average of the sum of the conductance values in the form of the histogram are shown in the inset in panel (e). The histograms peaks at 0.75, and the average conductance spatially varies in a range about 30 percent to that of the full range. In particular 5 percent change at the right edge of the histogram pertains to the slight contrast present between the regions with two distinct conductance behavior shown in panel (d). The other observation from panel (e) is that the average conductance deviates (in fact, is lower) from unity which would be in the case of a BCS-superconducting gap. In panel (f), the underlying image of the selected surface is shown. The color changes represent the movement of the tip-position which is confined within 3 nm. However, these are not the changes of the topographic profile, but the effect of the tip-interaction at the positions of the strong charge density modulations present over the entire surface. In the central part of the image, the parallel lines are indicative of the beating positions of the charge density modulations. In some sense the panel (f) shows the density of the charge redistributed in the film in the presence of disorder. Clearly areas surrounding the beating positions are strongly affected compared to the regions slightly far. These areas where density of states also strongly fluctuates form zones, where due to a strong charge density, the superconducting gap does not develop. The absence of quasi-particle peaks assert that there is a tendency to phase break at these positions weaking the superconducting state.

In Figure 3, we provide the conductance maps at selected bias voltages from 0.0 mV to 0.75 mV at an interval of 0.05 mV. In each panel, the bias voltage is shown in the top left corner, and the color scale represents the change of the conductance pertaining to the maximal values at that bias voltage. Such a represenation is chosen in order to highlight visually the change in the spatial conductance maps with the increase in the bias voltage. At V = 0.0 mV, a small contrast is seen in the region which have the conductance curves without the quasi-particle peaks. We believe this should mimic the fluctuations in the density of states when the film is normal. The contrast between the two regions (with and without quasi-particle peaks) increases with the bias voltage, and at 0.25 mV, the maximum contrast can be seen in the conductance map. With further increase in the bias voltage, the contrast decreases, and nearly disappears at 0.4 mV, except at the interface between the regions separating two types of conductance behaviors, where the white lines appear which corresponds to the lower conductance values for these curves. The near uniform contrast at 0.4 mV corresponds to the crossover of the majority of the tunneling conductance curves exhibiting two behaviors. From 0.45 mV the bright regions highlight the reversal of the relative conductance changes between the two regions, and it continues to peak at the bias voltage of 0.55 mV, close to the position of the quasi-particle peaks in the BCS-like curves. Further increase in the bias voltage decreases the contrast, and near uniform conductance maps are seen from 0.7 mV onwards upto 1 mV. The uniform conductance (without any trace of two different regions) at 0.75 mV suggests that the spatial conductance changes occur only in the range of bias voltage, corresponding to the opening of the gap, asserting the locally varying electronic behavior in the film with the formation of the superconducting phase. Our measurements suggest that the effect of the disordered electron scattering on the superconducting phase is to produce concurrent conductance fluctuations at energies close to the superconducting energy gap at the beat positions of the charge density modulations. The zero field data imply that the superconducting conductance curves deviates from the s-wave BCS-like behavior, with reduced quasi-particle peaks. These are typical STS observations close to the superconductor to insulator transition in the disordered films. Therefore we believe the spatial changes in the local properties in the superconductins phase reflect a displacement of electronic states, and the total amount of states remains roughly the same in the normal state, at the energy range studied here. These observations may be extended to the other disordered superconducting films.

\paragraph{V-shaped tunneling density of states in superconducting and the normal state\\}

Further insight in the normal state can be obtained from the representative conductance curves in both the regions (with relatively homogeneous and inhomogeneous sueprconducting phase) at the selected magnetic fields, as shown in Figure 4. In panel (a), the curves with quasi-particle peaks shown in blue color are at zero magnetic field, and with the increase in the field to 1.5 T, the peaks are suppressed, and disappear at 4 T. In the regions where quasiparticle peaks are absent, the conductance curves merely widen at higher fields upto 4 T. The conductance curve at 4 T also near uniformly shows a considerable reduced states near zero bias voltage in this film. In order to compare the changes at zero magnetic field, with respect to the conductance at 4 T, the normalized curves are shown in the panels, (c) and (d), respectively for the representative curves in both the regions. With this procedure, the conductance curve at 4 T is brought to near unity at all the bias voltage values, and the relative differences are seen. At 1.5 T in panel (c) and (d), the increase in the conductance is about equal mangitude for both the regions. This observation points out that at 0 T, on the background of the nearly equal average conductance (i.e., at 4 T), the quasi-particle peaks develop with different heights, while forming the superconducting phase in this film. It also relates to the spatial map of the average conductance in Fig. 2(e), showing a slightly higher contrast in the central region, and a sharper edge in the histogram peak shown in its inset. The panel (e) in fig. 4 shows the conductance behavior at 4 T, in the bias voltage range from -60 mV to +60 mV, at T = 0.15 K. The tunneling conductance monotonically reduces as zero bias votage is approached, and the peculiar variation on the logarithmic scale, shown in the inset, has perfectly linear behavior. In panel (f) we show at the selected magnetic fields between 0 T to 5 T, the zero bias conductance in the curves at 100 mK. The initial increase in the zero bias conductance is gradual, whereas a saturation behavior is observed above 2.5 T. The slope of this curve is plotted in the inset, and a peak in the slope is seen, marked as $H_{max}$, at 2.5 T. This value is very close to the estimated $B_{c2}$ from the resistance measurements in TiN film suggesting that the conductance at 4 T, and onwards corresponds to the normal phase of the TiN film. Another significant observation that the variation of the conductance is logarithmic relates to the long range Coulomb electron-electron repulsion being enhanced in the quasi-2D films of TiN due to the presence of disorder\cite{Altshuler1,Altshuler2,Bartosch,Kulkarni}. Our measurement asserts that the Cooper pairing emerges out of an electronic structure with a strongly reduced density of states at the Fermi level. Remarkably, the energy range for this reduction coincides with the superconducting gap. The reduced tunneling conductance features far from BCS s-wave theory can also be related to strongly reduced electronic states close to the Fermi level. We do not observe such visible spatial changes at high magnetic fields to be able to relate to the inhomogenieties in the superconducting phase. However, the specral map of the average conductance at 0 T (for e.g., panel (f) in Figure 2) indicates towards a reasonable variations in the density of states even in the normal state.

\paragraph{Discussions and Conclusions\\}

Though the microscopic origin of the disorder needs substantial studies of the sample properties at the nanoscale, the possible role of crystalline boundaries or the impurities from the precursors during sample preparation (chlorine impurities in the case of ALD deposited TiN films) as well as partial oxidation of the surface in the ambient atmosphere cannot be ruled out. Charge density modulations observed at the crystalline boundaries (Figure S2) suggest that the microscopic drive towards the disorder induced insulating like phase might set in locally by the inelastic scattering in some areas of the film. At the same time the diffusive motion of electrons sets up the long range correlation of the charge density modulations. Zero bias anomaly in the TiN film in the normal state relates to the disorder enhanced repulsive Coulomb interaction which leads to deviation of the BCS gap in the clean superconducting TiN film. Clearly determination of the superconducting gap is difficult in these films as the tunneling spectra contain information on both the pairing and repulsive interactions. The spatial inhomogeneities in the superconducting gap are therefore difficult to establish in the 2D disordered thin films. The spatial positions where the quasi-particle peaks are suppressed but possibly the superconducting gap still surviving may be interpreted as a result of the phase fluctuations. This indicates a possibility of localization of cooper pairs in these regions, amid a general bosonic character of the remaining of the film. With further increase in the disorder we expect the zero bias anomaly to pronounce and the superconducting gap will further reduce\cite{Kulkarni}. Close to the SIT, the Coulomb repulsion as well may become so strong that the percolation mechanism will cease to exist and the superconducting gap vanishes. The formation of the localized superconducting regions co-existing with the strongly disordered 2d-metallic phase across SIT may be a very remote possibility. The complete suppression of the superconducting state at SIT follows the loss of long range phase coherence in the entire film at a critical disorder and ascertains a fermionic mechanism. Till now there is no experimental evidence for the superconducting gap to be present in the insulating side or whether the spatial inhomogeneity on the superconducting side will lead to a Bose insulating phase across SIT.  However, in our results we find evidences of the regions of localized cooper pairs surrounded by the phase correlated superconducting areas. The interface of the two regions exhibit intermediate curves (at bias voltage 0.4 mV in Figure 3) with relatively energy independent density of states. Our results assert the possibility of superconducting and bose insulating regions co-exististing if the local disorder and the charge density modulations segregates two behaviors in the 2D films.
In conclusion the inhomogeneous superconducting phase was established in the disordered TiN film as a result of phase fluctuations and evidences are provided for the localized bose insulating regions in the superconducting film.

%%%%%%%%%%%%%%%%%%%%%%%%%
\begin{figure}
\includegraphics[width=0.9\textwidth]{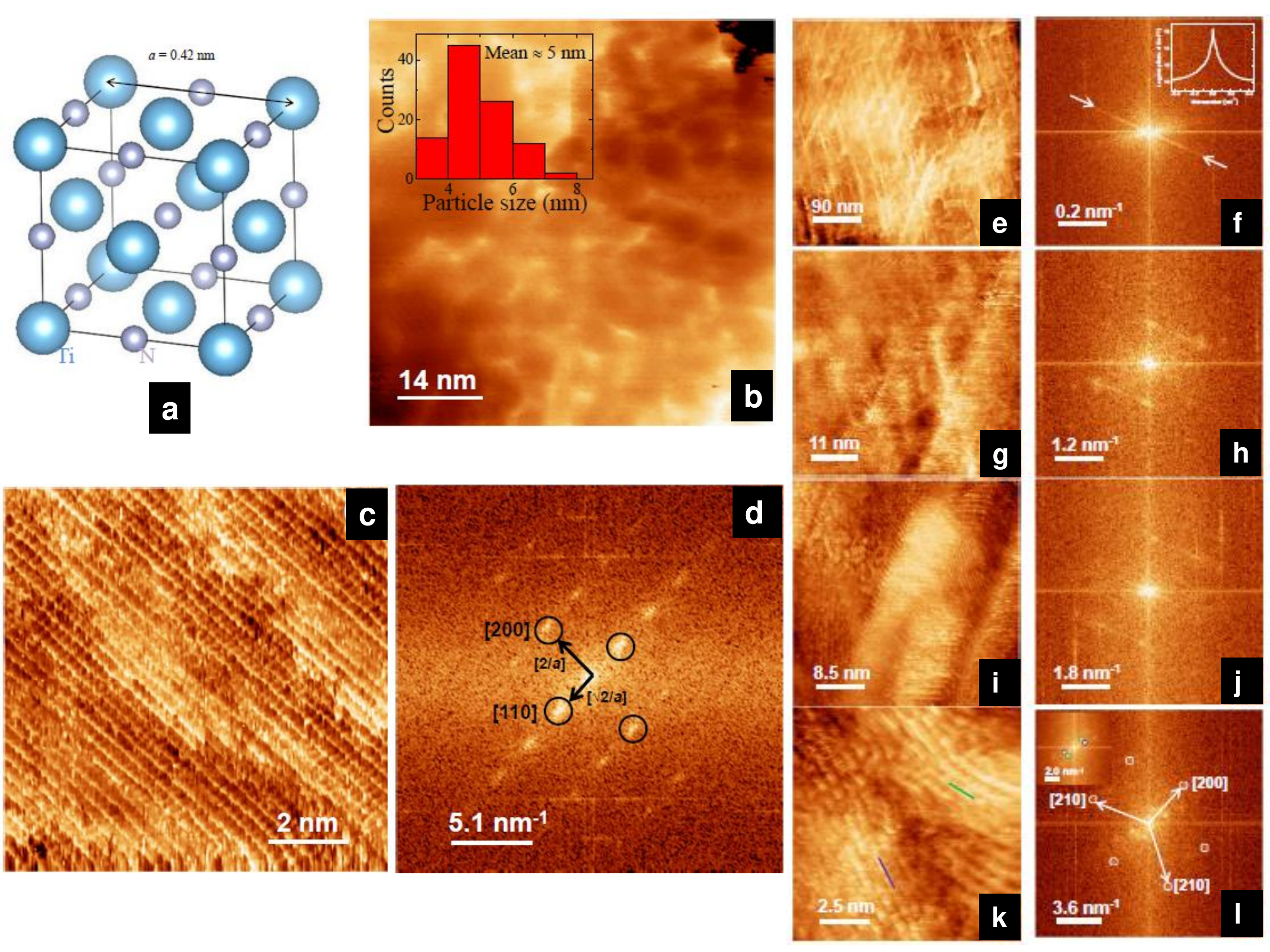}
\begin{center}
    \textbf{Figure 1}
\end{center}
\caption{\textbf{Atomic scale imaging in TiN films at 100 mK} (a) The crystal structure of titanium nitride with Ti atoms forming the FCC unit cell, and the N atoms occupying the positions at the centre, and on the edges. The lattice parameter for [100] plane is shown as 0.42 nm. The panel (b) shows the surface image for the TiN film. The crystallite sizes were estimated and presented in the form of the histogram shown in the inset. \textbf{c} STM topographic image over an area 10 x 10 nm taken at $100$mK in the superconducting TiN thin film. The corresponding 2d-FFT image is shown in panel (d). The first set of four bright spots are encircled using black color line. Arrows show the distances in the units of lattice parameter $a$ and the corresponding the planes are labeled in the square brackets. Panels(e,g,i,k) show Charge density modulations and their patterns over large to smaller areas. In panels (f,h,j,l) are corresponding 2d-FFT images. Inset in panel f shows the line profile across the white arrows in the main panel. In panel l, the two sets of bright spots are encircled, one in the main panel and other in the inset. The corresponding set of planes are indexed in the main panel (l). The two different orientations of charge density modulations are identified as the green and purple line in panel k, and the corresponding bright spots in the 2d-FFT are in the inset in panel l.}
\end{figure}

\begin{figure}
\includegraphics[width=0.9\textwidth]{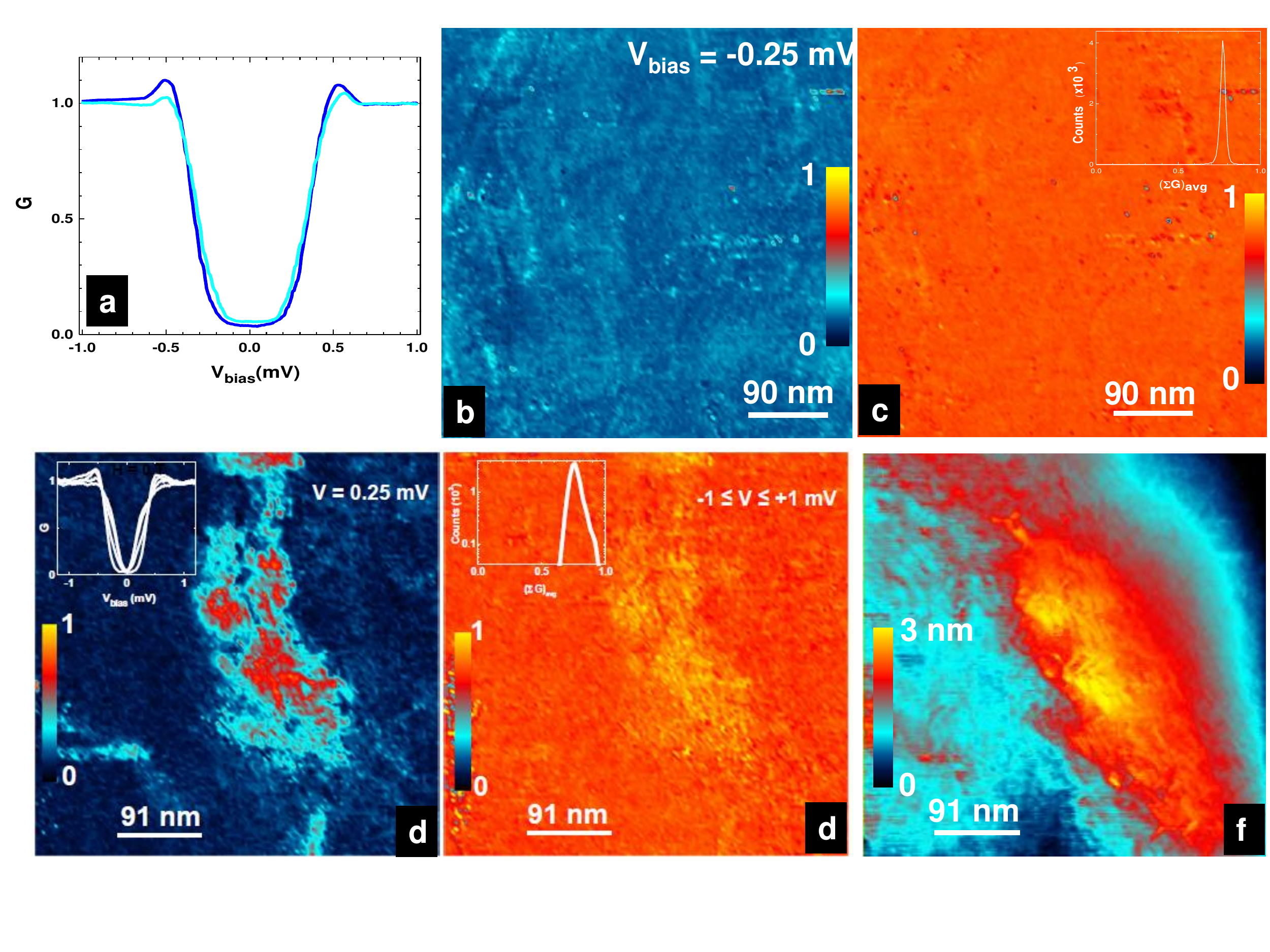}
\begin{center}
    \textbf{Figure 2}
\end{center}
\caption{\textbf{STS image on the Charge density modulations and their patterns} Panel (a) shows the representative tunneling conductance curves, and (b) shows the spectromap at the bias voltage of -0.25 mV. The map comprises 128 x 128 values spatially distributed over an area 512 x 512 nm. Panel (c) shows the average conductance on the same area, with the inset showing the histogram of the average condutance for each curve. Main panel (d) in the lower set of figures shows the conductance distribution at V$_{bias}$ = 0.25 mV spread over an area 455 nm x 455 nm measured at zero magnetic field. The color scale represents the normalized conductance with dark blue regions in the map have near zero values and the red colored regions showing conductance upto 0.75, i.e, 75 percent of the normalized conductance at unity. The representative conductance curves are shown in the inset in panel (d). The curve displaying the clear quasi-particle peaks is measured at dark blue regions in the main panel (d), and the curve with monotonic decrease of conductance as the bias voltage approaches zero mV is the representative of the red regions in the map. Two intermediate curves measured in the vicinity of these two regions are also shown in the inset. In panel (f) the color scale represents the average of the sum of the conductance for each curve, in the bias voltage range from -1 to +1 mV. In the inset in panel (f), the histogram of the average of the sum of the conductance, with 128 x 128 values, is shown and their spatial map is shown in the main panel (f).}
\end{figure}

\begin{figure}
\includegraphics[width=0.9\textwidth]{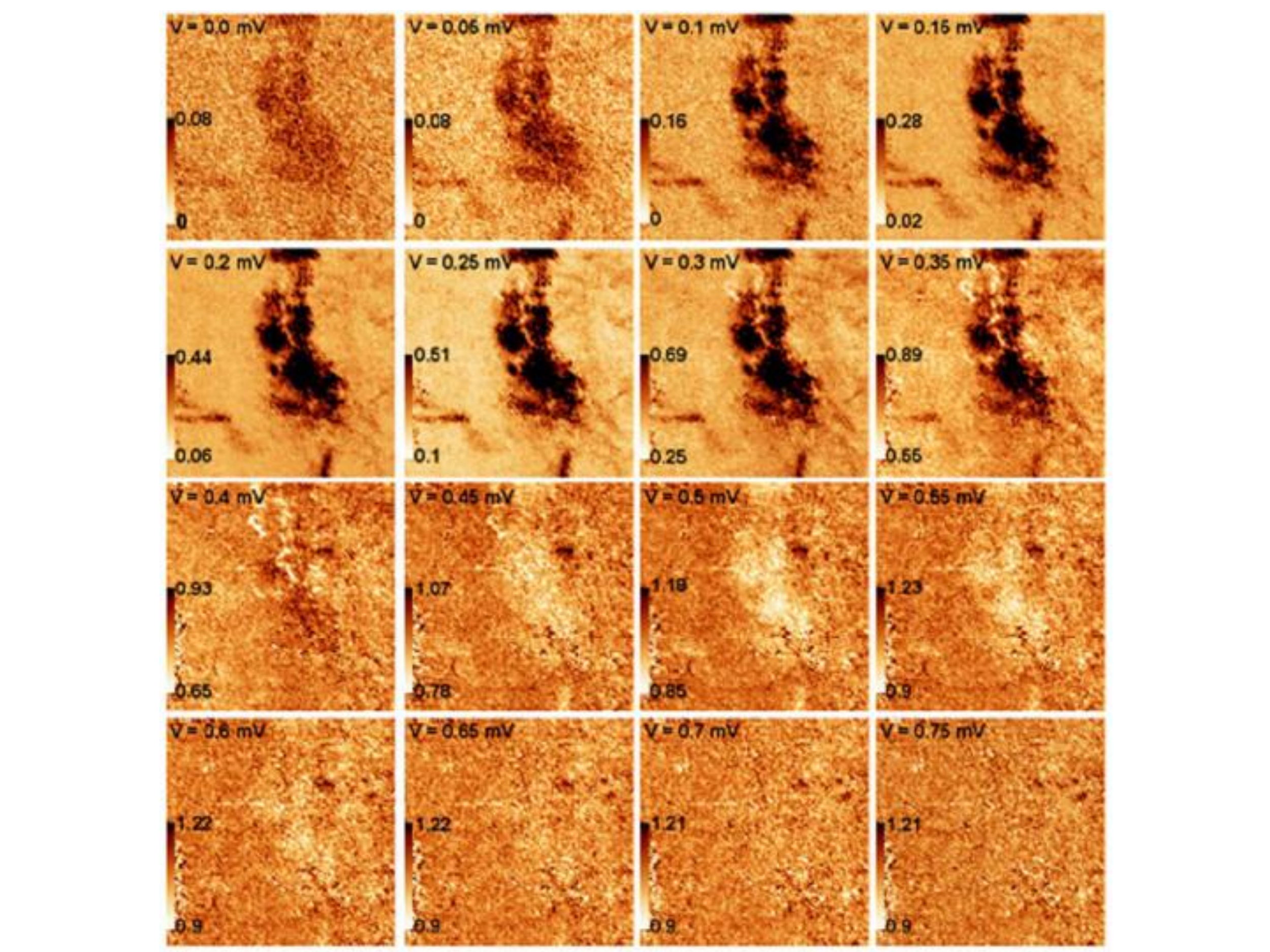}
\begin{center}
    \textbf{Figure 3}
\end{center}
\caption{\textbf{Bias voltage dependence of the inhomogeneous superconducting state}. The conductance maps made at 0.05 mV interval in the bias voltage range from 0 to 0.75 mV. The labels in the top left corner in each panel show the bias voltage for each conductance map. The scale in the bottom left corner shows the range of the conductance which is kept variable in order to show the maps with the same contrast. The white regions are with the highest tunneling conductance for the given bias voltage and the black regions are near zero values. The maximum contrast is at V$_{bias}$ = 0.25 and 0.5 mV. These are positions with the maximum changes in the conductance for the fixed bias voltage. The bias voltage 0.4 mV corresponds to the crossover of the two sets of tunneling conductance curves. As such there is no variance of the tunneling conductance at 0.4 mV except for the bright lines appearing at the interface regions which separates the two behaviors of the tunneling conductance curves. The appearance of these lines start at 0.3 mV and persists till at 0.45 mV.}
\end{figure}

\begin{figure}
\includegraphics[width=0.9\textwidth]{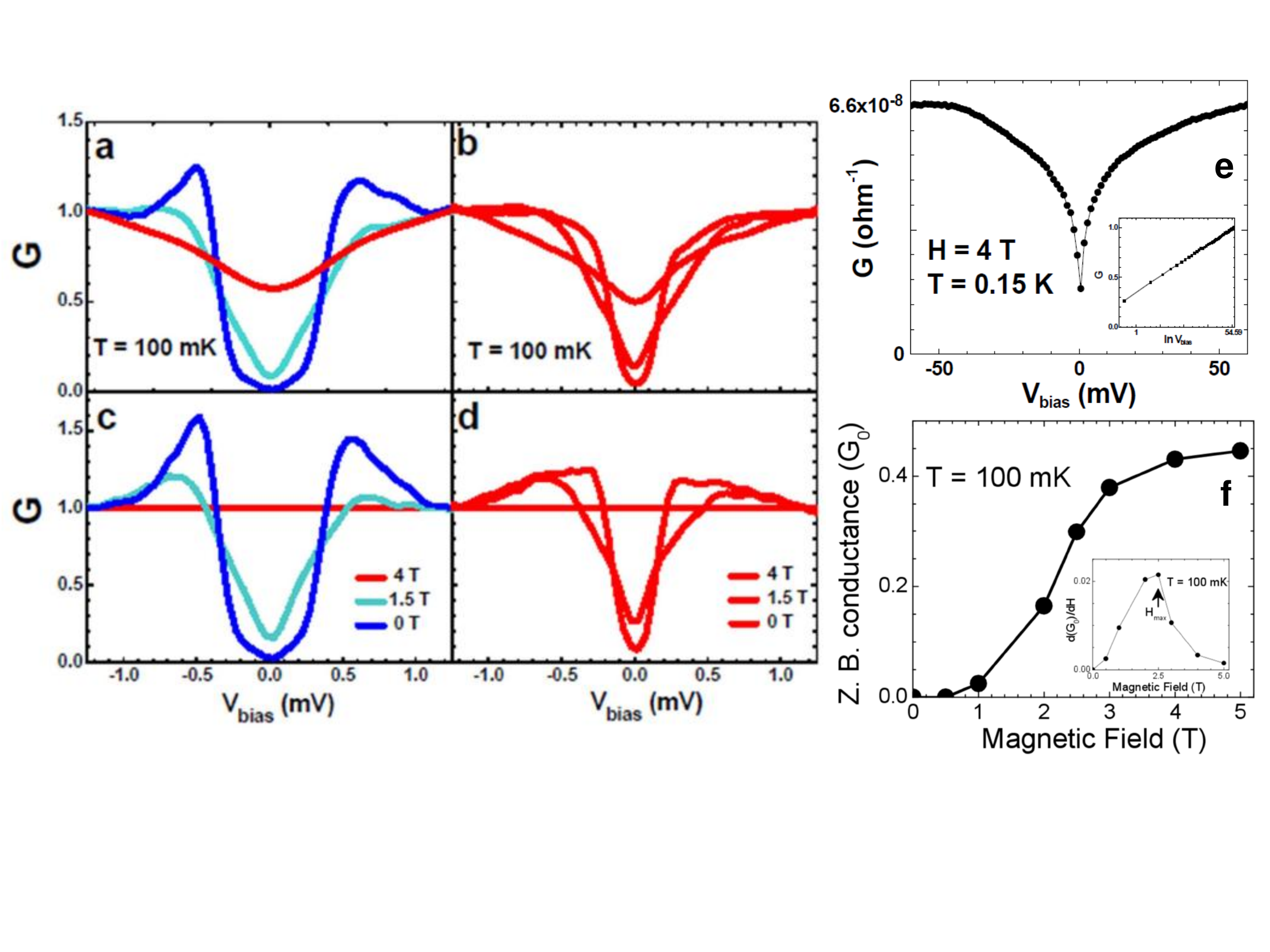}
\begin{center}
    \textbf{Figure 4}
\end{center}
\caption{\textbf{Comparison of the conductance curves at selected magnetic fields.} Panels (a) and (b) show the conductance curves at selected fields in dark blue and red color regions, respectively, in Fig. 2(d), i.e., both the regions. In panels (c) and (d) we plot the conductance at 4 T as unity at all bias voltages and renormalize the conductance at 0 and 1.5 T with respect to the 4 T curve.This is done by a numerical subtraction of the tunneling conductance curves for both the regions. In the main panel (e) the conductance curve at T = 150 mK is shown in the bais voltage range from -60 to + 60 mV. The curve corresponds to the magnetic field H = 4 T, and in the inset the same curve is shown in the logarithmic scale of the bias voltage. In the main panel (f) we plot the zero bias conductance values as a function of applied magnetic field at T = 100 mK. Note the initial rapid increase in the zero bias conductance, and the saturation behavior for higher fields. This is represented using a derivative of the curve is shown in the inset, with the peak in the derivative marked by an arrow as $H_{max}$ at 2.5 T. This is close to the estimated value of the upper critical field using the transport measurements in the TiN film.}
\end{figure}

\begin{addendum}
 \item The Laboratorio de Bajas Temperaturas is associated to the ICMM of the CSIC. This work was supported by the Spanish MICINN and MEC (Consolider Ingenio Molecular Nanoscience CSD2007-00010 program, FIS2011-23488, ACI-2009-0905 and FPU grant) and by the Comunidad de Madrid through program Nanobiomagnet.

 \item[Competing Interests] The authors declare that they have no competing financial interests.
 \item[Correspondence] Correspondence and Requests for materials should be addressed to P.K., prasanna1609@gmail.com.
\end{addendum}

\newpage
\begin{center}
    \textbf{Supplementary Information}
\end{center}

\section{Experimental arrangements and measurement of thin films with STM.}

We used the conventional STM set-up in a dilution refrigerator with the possibility to apply a magnetic field upto 8 T. Previous experiments in the same set-up have demonstrated subatomic spatial scanning capabilities, an energy resolution of 15 $\mu$eV in spectroscopy\cite{Crespo06a}, and a lowest temperature of 100 mK. To measure the TiN thin films, we mounted one thin film whose resistance vs temperature, current and magnetic field had been measured previously in Refs.\cite{Baturina}. We placed a mechanically sharpened tip of Au on top of the sample and cooled down. Close to the sample, we placed an Au pad to clean the tip. After cleaning the tip immediately on cool down, we moved in-situ the tip to the sample and obtained excellent tunneling conditions with large scale images reproducible as a function of the tunneling conductance and high valued work functions (several eV). Images show a flat surface with a small corrugation, below the nm.

\section{Electronic properties of TiN.}

Bulk TiN has a Fermi surface made out of three bands from bonding-antibonding mixtures of $2p-3d$ character crossing the Fermi level \cite{Haviland85,Winzer84,Ern65}. One leads to a cross centered aroung $\Gamma$ touching the Brillouin zone at the X points, an the other two are small electron pockets at $\Gamma$. Effective masses range from 1.1m$_0$ to 1.8m$_0$ (with m$_0$ the free electron mass), and the Fermi level is at a rising portion of the overall density of states, having close to it (at around 1 eV) a pronounced maximum. With disorder, the properties of the crystal are washed out, leading gradually to isotropic Fermi surfaces and a featureless density of states. Because the density of states is so close to a maximum, we can expect an increase in the effective mass of the electrons with increased scattering.

In our samples, being disordered and closer to the insulating phase, with a strong granular structure and a short mean free path, the concept of a Fermi surface at a macroscopic level is not well defined. From the resistivity, and the value of the superconducting coherence length ($\xi_0=9.5$ nm from B$_{c2}$), we can expect k$_F\ell\approx$1.5 (using an effective mass of m$^*=2m_0$) in the superconducting sample. However, within a single crystallite, electrons should behave qualitatively with properties similar to that in bulk single crystals, with a reduced mean free path. Thus, the size of the charge density modulations observed in the samples provides fixed points for electron ordering at the Fermi surface. The Bragg peak positions give the corresponding charge modulation wavevectors.

\section{Properties of TiN thin films.}

The TiN thin films were grown using atomic layer deposition using reactant gases TiCl$_4$ and NH$_3$\cite{Uhm01}. The growth of crystals is columnar-like with extremely small boundary regions between crystallites, ranging only of a few atomic layers\cite{Uhm01}. This aspect is confirmed in our atomic scale images (Fig S1). Impurities in the crystallites can be Carbon atoms and Chlorine, introduced during deposition. These are marginally present and are difficult to detect using surface analysis procedures. The nominal composition of our sample is TiN$_{0.94}$ with 3.5\% of Cl. One possibility is that Carbon and Chlorine defects are distributed over crystallite boundaries, producing electron scattering. The TiN films are very thin (5 nm is just some ten unit cells) and it can seem surprising that they remain intact for a long time after fabrication. This implies an extreme stability to chemical activity at ambient conditions, and has been studied previously, regarding the applications of this material e.g. in junction fabrication for microelectronic devices, cosmetics or optical filering \cite{Kim09,Wittner81,Zega77}. TiN adds up indeed the advantages of metallic Ti-Ti bonding for optics and the high melting point and hardness for applications requiring mechanical and thermal stability. TiN films have many applications, such as improving oxydation resistance in cutting tools\cite{Escobar05}. It is a truly stable material, as confirmed additionally in the present work through what we believe could be possibly the first STM observation of atomic size features in polycrystalline films.

\section{Charge density modulations over the atomic lattice}

In Figure S1 a and c, we show topographic images over 15 nm x 15 nm and 16.6 nm x 16.6 nm, respectively. The numerals in the image, 1 to 3, are representatives of three regions. 1 and 3 correspond to the two crystallites, which are seen with different atomic orientations. The boundary between these two regions is marked by numeral 2, where formation of charge density modulations can be seen. In the panel (b) and (d), the 2d-FFT image of panel (a and (c)), three configurations of the bright spots are seen. The encircled green spots represent the atomic orienations in the crystallite, marked by 3, in panel (a) and (c). The encircled red spots are FFT of the crystallite, marked by 1. A change of atomic orienations from 3 to 1 is about 30 degrees. The light blue spots with longer wavelengths represent the periodicity of the charge density modulations, which are seen confined at the crystalline boundary. This highlights the role of crystalline boundaries in disorder electron scattering and possibly the formation of standing wave patterns on the film surface. Such electron scattering at the crystalline bounaries may also result in the enhanced electron-elctron repulsion effect \cite{Altshuler1,Altshuler2,Bartosch}, in the 2-d disordered films\cite{Kulkarni}. 

In figure S2 (a) and (b), we show the filtered images of the charge modulations over atomic lattice. These images correpond to the charge density modulations seen in Figure 1(k). The 2d-FFT in Figure 1(l) was used to filter higher intensities, in order to bring out the underlying atomic level features. The panel (a) in the figure S2 shows the charge modulations as the bright waves running parallel in different orientaions. In the panel (b), the charge density modulations are seen regions overriding the underlying atomic lattice. The wavy pattern of the larger periodicity can be easily distinguished on the underlying rectangular lattice. 

\section{Superconducting state over the charge density modulations}

In figure S3, we show the STM/S measurement over an area 5 x 5 nm. Panel (c) shows the STM image of the underlying surface, where clearly visible are the atomic rows, and the co-existing charge density modulations. The inset in panel (c) shows the 2d-FFT image, where we encircle the bright intensities corresponding to the atomic lattice as well as the charge density modulations at higher wavelength. STS measurements over this region with 128 x 128 IV-curves were made. The representative IV-curves are shown in Panel (a), where the height of the quasi-particle peak fluctuates. The arrows tentatively show the range of fluctuations at -0.5 mV and -0.2 mV. The conductance distribution at these two bias voltages is shown as histogram in the inset in panel (b). The two peak positions show the most probable conductance and the width of the peak indicates the conductance distribution. The average conductance for each of the 128 x 128 curves was calculated, and plotted as a histogram, shown in the main panel (b). The average conductance peak is positioned at 0.82, with a small width of the distribution, lying below unity. The STS maps were made between -1.2 mV to +1.2 mV after the tunneling curves were normalized at 1 mV. In panel (d) at the selected bias voltage, V = -0.5 mV. The map is a visual of the the spatial conductance distribution at the bias voltage mentioned in the inset of panel (b). A 2d-FFT image of the conductance map was obtained, which is shown in the inset in panel (d). The bright intensities are encircled, in order to enhance the resembalance with the inset in panel (c). A reverse FFT which filters all other intensities recovers the underlying spectrograph, which clearly shows the atmoic level features in the superconducting state. The height of the coherence peaks fluctuates at the positions of the charge density modulations, and these are seen clearly in panel (d). Our measurements suggest that the effect of the disorder electron scattering on the superconducting phase is to produce concurrent conductance fluctuations at energies close to the superconducting energy gap. 

In figure S4 we show the conductance maps at three selected magnetic fields, H = 0 T, 1.5 T and 4 T, respectively in the panels (a-c). The conductance maps are plotted at a bias voltage 0.25 mV, with the curves having been normalized at 1 mV. The area under the investigation is 950 nm x 950 nm, and is the same at three selected fields, except for a slight shift of the order of few nanometers at higher fields. The color scale in the panel (c) is also the same as that in the panel (a) in Figure 2 in the main text. The regions in the red color in the panel Fig. S4(a) are those with the conductance curves without quasi-particle peaks at zero magnetic field. The contrast between two regions reduces with the increase in the magnetic field to 1.5 T. The conductance map at 4 T does not show any contrast between the two regions, instead near uniform conductace values are spread over the region where originally a strong change in the contrast was observed at 0 T. At the outset such dramatic change in the conductance maps with application of the magnetic field asserts that the inhomogeneinities at 0 T are resulting from the local electronic properties of the superconducting phase.

\newpage 

\begin{figure}
\includegraphics[width=0.9\textwidth]{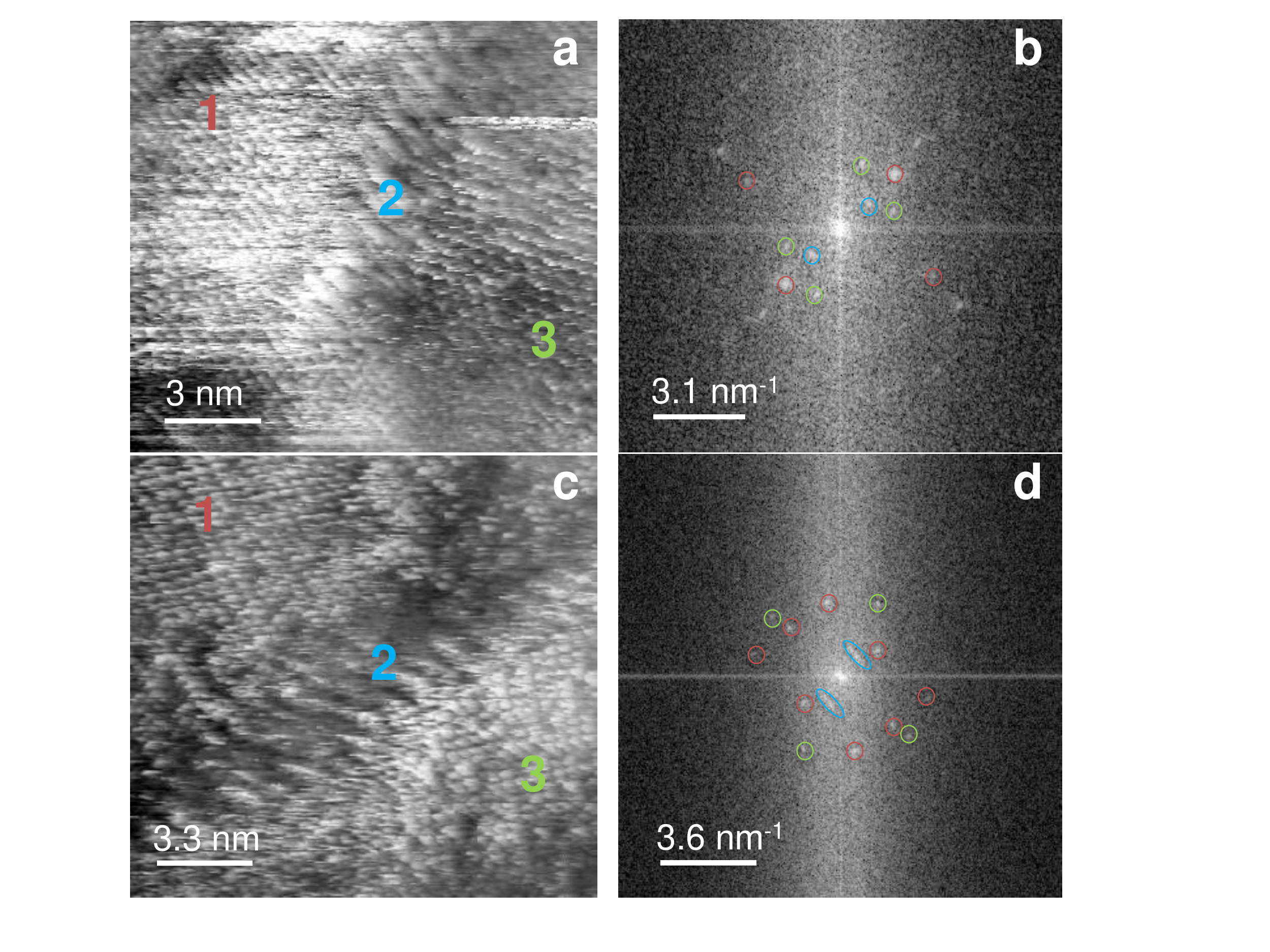}
\begin{center}
    \textbf{Figure S1}
\end{center}
\caption{\textbf{S1 Polycrystalline imaging using STM} Panel (a) shows STM image over 15 x 15 and (c) 16 nm x 16 nm at 100 mK and at zero magnetic field and at 4 T respectively. The 2d-FFT image is shown in panel (b) and panel (d), where several bright spots are encircled with three different colors. The red and green correspond to regions on the top of the crystallites, and the inner blue circles show spots on the crystalline boudaries where charge density modulations are observed.}
\end{figure}

\begin{figure}
\includegraphics[width=0.9\textwidth]{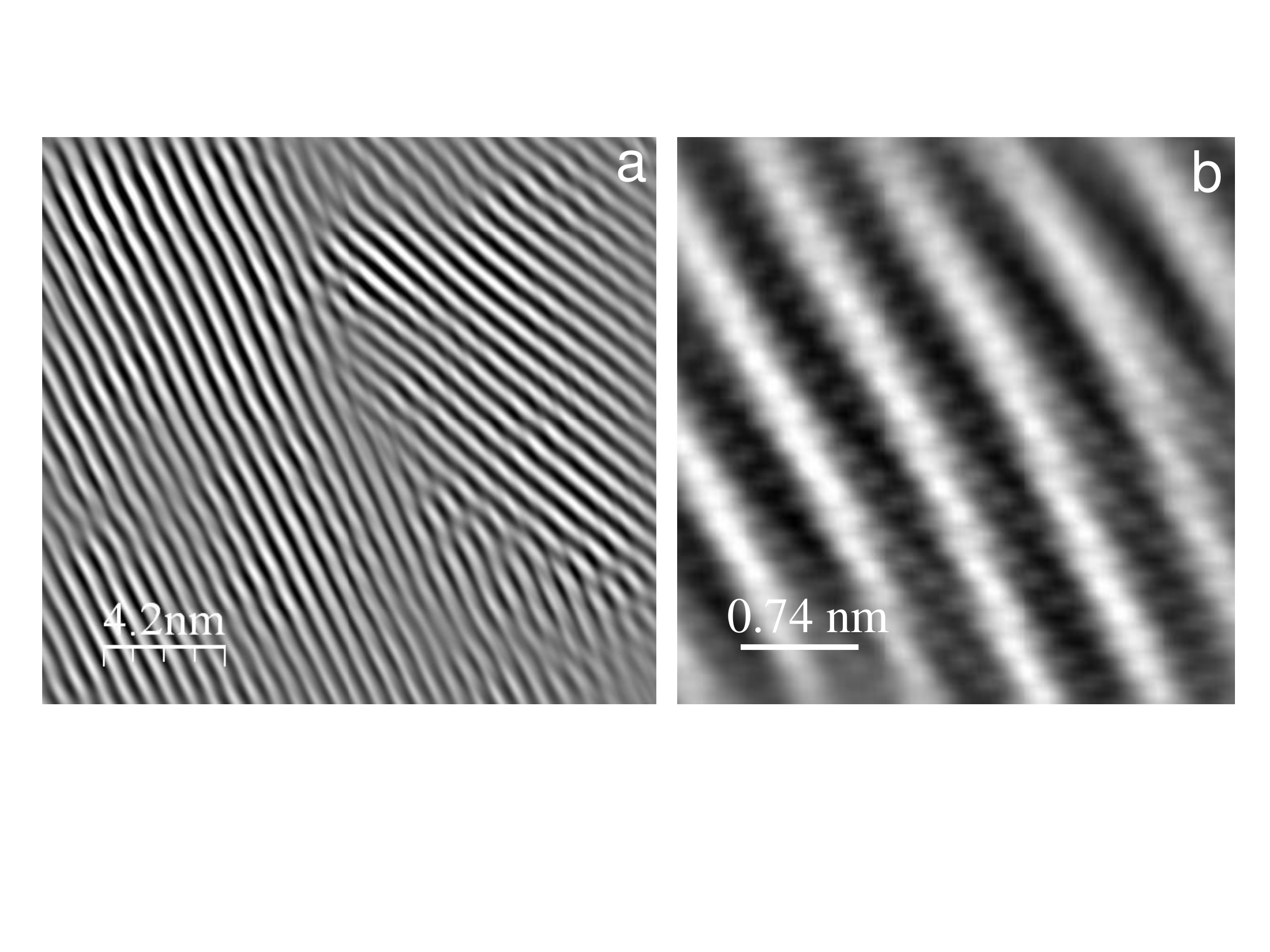}
\begin{center}
    \textbf{Figure S2}
\end{center}
\caption{\textbf{S2 Changes in orientations of the charge density modulations over nanometer scale areas.} In the left panel of (a) we show the filtered image of the charge density modulations seen in panel (k) in Main figure 1. The bright spots in the inset in panel Figure1 (l) were used to filter all other intensities from the real space image of the charge density modulations only. In panel S2 (b), the zoom up of a small region is shown, which was obtained using the bright spots in the main panel (l) in Figure 1. In panel S2 (b) the charge density modultions are seen riding on the underlying atomic lattice.}
\end{figure}

\begin{figure}
\includegraphics[width=0.9\textwidth]{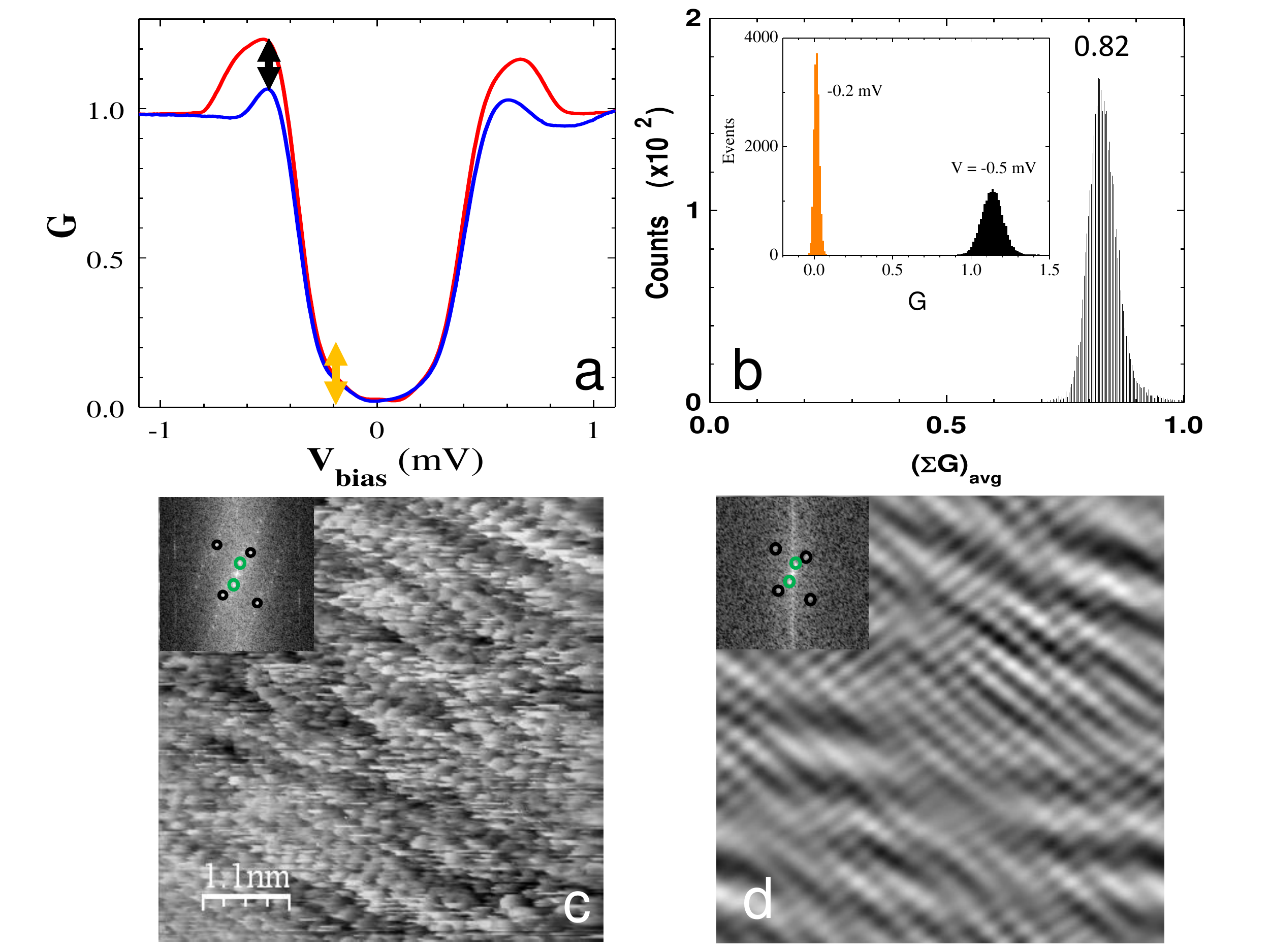}
\begin{center}
    \textbf{Figure S3}
\end{center}
\caption{\textbf{S3 Atomic level features of the Superconducting state.} Here we show in panel (a) the representative tunneling conductance curves measured on the 5 x 5 nm surface shown in panel(c). In panel (a) the black and the yellow arrow shows the bias voltage position of -0.5 mV and -0.2 mV, respectively. The range of conductance values at these bias voltages are quantified and plotted as histograms, as shown in the inset in panel (c). The main panel shows the histogram of the average conductance, calculated from each of the 128 x 128 tunneling curve. The inset in panel (c) shows the 2d-FFT of the surface image of the main panel. The inner green circles contain the bright spots corresponding to longer wavelength of the charge density modulations, while the outer black circles enclose the bright spots of the underlying atomic lattice. The inset in panel (d) shows the 2d-FFT image of the spectromap made at -0.5 mV over the surace in panel (c). The filtered image is shown in main panel (d).}
\end{figure}

\begin{figure}
\includegraphics[width=0.9\textwidth]{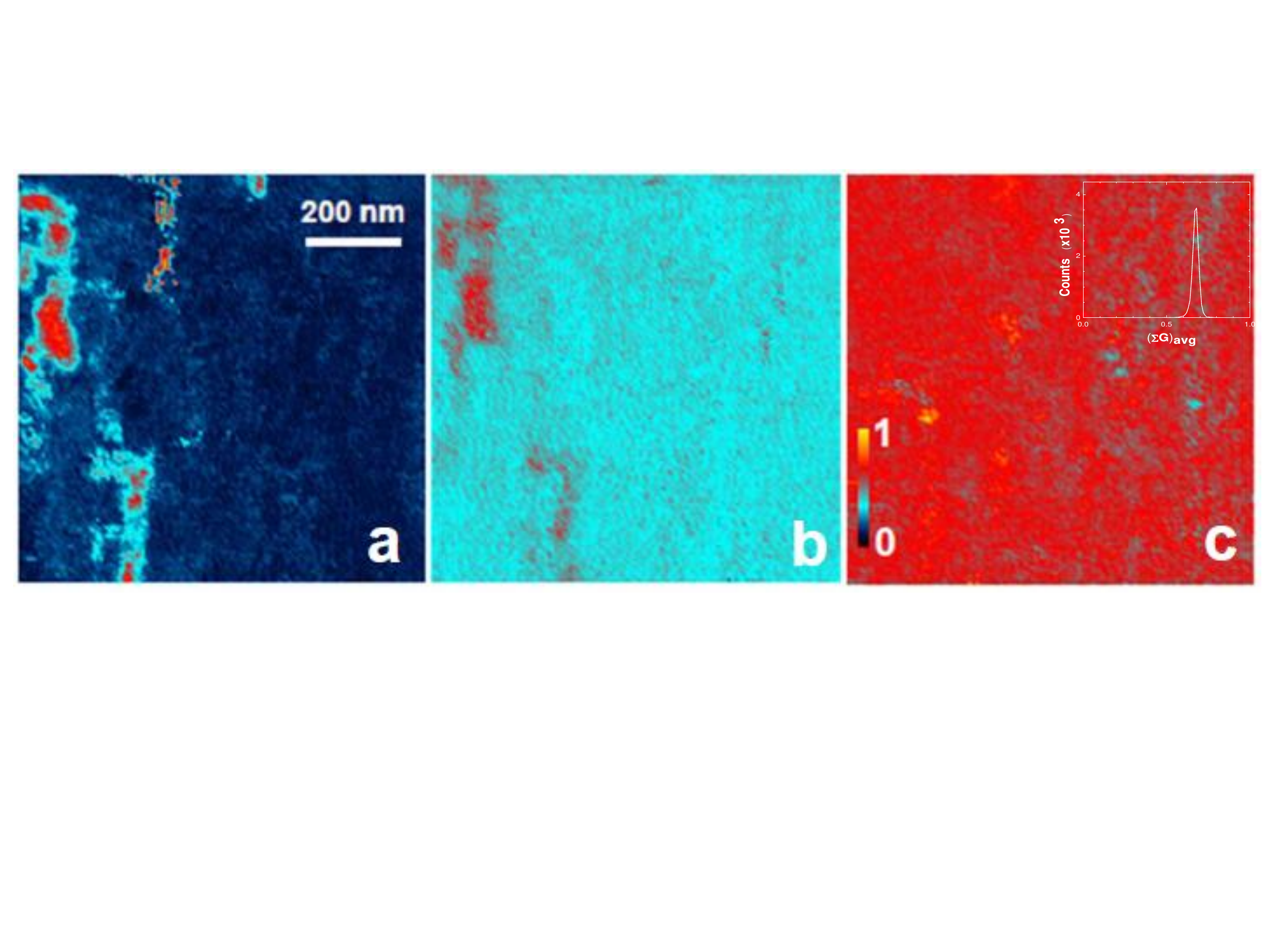}
\begin{center}
    \textbf{Figure S4}
\end{center}
\caption{\textbf{S4 Magnetic field dependence of the inhomogeneous superconducting state.} In the main panel (a) the white color bar at the top right corner is the length scale of the area 950 nm x 950 nm. The panel (a) is the conductance map at $V_{bias}$ = 0.25 mV at H = 0 T. The color changes correspond to the conductance values following the scale shown in the panel (c), which shows the map at H = 4 T. Total 128 x 128 conductance values are plotted over a normalized scale of 0 to 1 and it is also common to the conductance map at H = 1.5 T shown in panel (b). Inset in panel (c) shows the sum of the average conductance for all three magnetic fields, estimated in the bias voltage range from -1 to +1 mV. The average conductance remain approximately same in all three cases.}
\end{figure}

\end{document}